# Detecting Effects of Filaments on Galaxy Properties in the Sloan Digital Sky Survey III

Yen-Chi Chen,[1][⋆] Shirley Ho,[2,3] Rachel Mandelbaum,[3,4] Neta A. Bahcall,[5] Joel R. Brownstein,[6] Peter E. Freeman,[4,7] Christopher R. Genovese,[4,7] Donald P. Schneider,[8,9] Larry Wasserman[4,7]

[1]*Department of Statistics, University of Washington, Seattle, WA 98195, USA*
[2]*Lawrence Berkeley National Lab, Berkeley, CA 94720, USA*
[3]*Department of Physics, Carnegie Mellon University, Pittsburgh, PA 15213, USA*
[4]*McWilliams Center for Cosmology, Carnegie Mellon University, Pittsburgh, PA 15213, USA*
[5]*Department of Astrophysical Sciences, Princeton University, Princeton, NJ, 08544, USA*
[6]*Department of Physics and Astronomy, University of Utah, 115 S 1400 E, Salt Lake City, UT 84112, USA*
[7]*Department of Statistics, Carnegie Mellon University, Pittsburgh, PA 15213, USA*
[8]*Department of Astronomy and Astrophysics, The Pennsylvania State University, University Park, PA 16802, USA*
[9]*Institute for Gravitation and the Cosmos, The Pennsylvania State University, University Park, PA 16802, USA*

28 November 2016

**ABSTRACT**
We study the effects of filaments on galaxy properties in the Sloan Digital Sky Survey (SDSS) Data Release 12 using filaments from the 'Cosmic Web Reconstruction' catalogue (Chen et al. 2015a), a publicly available filament catalogue for SDSS. Since filaments are tracers of medium-to-high density regions, we expect that galaxy properties associated with the environment are dependent on the distance to the nearest filament. Our analysis demonstrates that a red galaxy or a high-mass galaxy tend to reside closer to filaments than a blue or low-mass galaxy. After adjusting the effect from stellar mass, on average, early-forming galaxies or large galaxies have a shorter distance to filaments than late-forming galaxies or small galaxies. For the Main galaxy sample (MGS), all signals are very significant ($> 6\sigma$). For the LOWZ and CMASS sample, the stellar mass and size are significant ($> 2\sigma$). The filament effects we observe persist until $z = 0.7$ (the edge of the CMASS sample). Comparing our results to those using the galaxy distances from redMaPPer galaxy clusters as a reference, we find a similar result between filaments and clusters. Moreover, we find that the effect of clusters on the stellar mass of nearby galaxies depends on the galaxy's filamentary environment. Our findings illustrate the strong correlation of galaxy properties with proximity to density ridges, strongly supporting the claim that density ridges are good tracers of filaments.

## 1 INTRODUCTION

Matter in our Universe tends to aggregate around certain low-dimensional structures that form the Universe into a network called the cosmic web (Bond et al. 1996). The early work on the evolution of cosmic web and its relation to the galaxy formation can be dated to the 70s (Doroshkevich 1970; Zeldovich & Novikov 1975; Doroshkevich et al. 1980; Zeldovich et al. 1982a; Zeldovich 1982). The cosmic web consists of four distinct types of sub-structures: highly concentrated clusters, elongated filaments, widely spread sheets, and voids (Zeldovich et al. 1982a). In this study, we focus on filaments for several reasons. First, a large fraction of the matter in the Universe is contained in and around filaments (Aragón-Calvo et al. 2010; Eardley et al. 2015), allowing detection of the correlation between filaments and properties of nearby galaxies even when the correlation is weak. Second, filaments are similar to clusters in the sense that they both occupy higher density regions (in comparison to walls and voids). We therefore expect galaxies close to filaments to share some characteristics of galaxies around clusters. Moreover, filaments are where the matter caustics occur so they represent a special regime within the large-scale structure. Lastly, relatively few studies have been performed for filaments compared to clusters (some recent work can be found in Tempel et al. 2014b; Guo et al. 2015; Tempel et al. 2015; Zhang et al. 2015).

Despite the vagueness of the formal definition of filaments, they are typically described as curve-like tracers of high-density

[⋆] E-mail: yenchic@uw.edu

regions of the Universe (Bond et al. 1996). As studies of galaxy clusters have shown, galaxy properties are dependent on the density of the environment (Butcher & Oemler 1978; Bower et al. 1992). We are therefore interested whether similar trends exist for galaxies in or around filaments.

Vast amounts of evidence demonstrate that the environment around a galaxy affects that galaxy's star formation; see, e.g., Kauffmann et al. (2004); Blanton et al. (2005b); Christlein & Zabludoff (2005); González & Padilla (2009); Creasey et al. (2015). Moreover, direct evidence also suggests that star formation is related to the nearby filaments (Darvish et al. 2014). Hence, galaxy properties related to star formation, such as the color, number of satellites, stellar mass and disk galaxy spin alignment (Robertson et al. 2005; Lagos et al. 2009; González & Padilla 2009; Guo et al. 2015; Codis et al. 2015), are expected to be influenced by the environment. In particular, direct evidence has indicated that a galaxy's color is generally correlated with the environment; see, e.g., Hogg et al. (2003); Balogh et al. (2004); Springel et al. (2005); Park et al. (2007); Coil et al. (2008); Font et al. (2008); Guo et al. (2011). These findings suggest that red galaxies tend to reside in high-density regions while blue galaxies tend to live in low-density regions (Cowan & Ivezić 2008; Grützbauch et al. 2011).

In addition to the color, other quantities such as stellar mass, size and age of a galaxy are related to the environment. Grützbauch et al. (2011) found that stellar mass is correlated with the local density, i.e., galaxies located in high-density regions tend to be more massive. Moreover, direct evidence reported by the GAMA (Galaxy And Mass Assembly) survey has shown that galaxies within different environments have different stellar mass distributions (Alpaslan et al. 2015). Indirect evidence also links the stellar mass to the environment by the stellar mass-halo mass ratio (Moster et al. 2010) and the fact that environment impacts halo formation (Desjacques 2008). The size-environment relation has been observed in Cooper et al. (2012) and Lani et al. (2013), and the environment is also correlated with the Fundamental Plane relating velocity dispersion, surface brightness, and size of elliptical galaxies (Joachimi et al. 2015). Moreover, the size-stellar mass relation is dependent on the environment (Cappellari 2013; Kelkar et al. 2015). In addition to stellar mass and size, many studies have shown that the age of a galaxy is environment-dependent; see e.g. (Bernardi et al. 1998; Trager et al. 2000; Sil'chenko 2006; Bernardi et al. 2006; Wegner & Grogin 2008; Smith et al. 2012a; Deng 2014). Since filaments are tracers of medium-to-high density regions, we expect that all these galaxy properties are correlated with proximity to filaments as well.

Besides the above galaxy properties, due to the tidal and velocity field around filaments (Hahn et al. 2007a,b; Tempel et al. 2014a), spin and principal axes of a galaxy are known to be correlated with orientation of nearby filaments, see, e.g., Altay et al. (2006); Tempel et al. (2013); Tempel & Libeskind (2013); Aragon-Calvo & Yang (2014); Dubois et al. (2014); Chen et al. (2015b). However, in the current paper, we focus only on the the color, stellar mass, age, and size of a galaxy and study how these properties may be related to the distance to filaments.

The fact that galaxies around filaments and clusters share some characteristics can be used to test the consistency of a filament finding technique. An issue for filament detection is that there is no consensus on the precise definition of filaments. There is only a general, qualitative agreement that they are curve-like structures that trace high-density regions (Bond et al. 1996). The term highdensity is used here is to compare with cosmic sheets or voids. Most of the current state-of-the-art filament finders, such as the Multi-scale Morphology Filter (MMF; Aragón-Calvo et al. 2007, 2010), the NEXUS and NEXUS+ (Cautun et al. 2013), the Candy model (Stoica et al. 2007; Stoica et al. 2005), the skeleton (Novikov et al. 2006; Sousbie et al. 2008a,b), and the DisPerSE models (Sousbie 2011), all output filaments consistent with this high-density property. Comparing a new filament finder to the existing ones may not be an optimal way to check the consistency of detecting real filaments; a better approach is to compare the properties of those galaxies that are around the filaments since, in theory, these galaxies should be similar to those close to clusters.

In this paper, we study properties of Sloan Digital Sky Survey (SDSS York et al. 2000; Eisenstein et al. 2011) galaxies around filamentsup to $z = 0.7$ using the 'Cosmic Web Reconstruction' filament catalogue (Chen et al. 2015a). Previous studies for filaments in SDSS mostly used the Main galaxy sample (Strauss et al. 2002) with $z \leqslant 0.25$ (Bond et al. 2010; Jasche et al. 2010; Smith et al. 2012b; Zhang et al. 2015; Leclercq et al. 2015). Generally, finding filaments beyond the Main galaxy sample catalogue is challenging due to the low observational number density, which decreases the detection accuracy drastically. By using density ridges as filaments (Chen et al. 2015c,a) and a process of redshift-slicing the Universe, the power of detecting filaments increases, which allows us to study the correlation between filaments and their nearby galaxies.

This paper is organized as follows. We begin with an introduction to the SDSS dataset in §2 and the filament catalogue in §3. We present our results for separating galaxies by the color (§4), stellar mass (§5), age (§6), and size (§7). Finally, we conclude this paper in §8.

We assume a WMAP7 ΛCDM cosmology with $H_0 = 70$, $\Omega_m = 0.274$, and $\Omega_\Lambda = 0.726$ (Anderson et al. 2012, 2014a).

## 2 THE SDSS DATA

We use three catalogs from SDSS data that contain the MGS sample from DR7 (Abazajian et al. 2009), and LOWZ and CMASS sample from DR12 (Alam et al. 2015).

SDSS I, II, and III together scanned 14,555 deg² of the sky using a five-band (*u, g, r, i, z*) photometric bandpasses (Fukugita et al. 1996; Doi et al. 2010) to a limiting magnitude of $r \simeq 22.5$. The resulting image data were then processed through a sequence of pipelines including astrometric calibration (Pier et al. 2003), photometric reduction (Lupton et al. 2001), and photometric calibration (Padmanabhan et al. 2008).

The SDSS DR7 (Abazajian et al. 2009) consists of the completed data set of SDSS-I and SDSS-II. These two surveys obtained wide-field CCD photometry (Gunn et al. 1998, 2006) in *u, g, r, i, z* photometric bandpasses (Fukugita et al. 1996; Doi et al. 2010), internally calibrated using the 'uber-calibration' process as described in Padmanabhan et al. (2008), forming a total footprint of 11,663 deg² of the sky. Based on the imaging data, galaxies within a region of 9380 deg² (Abazajian et al. 2009) were further selected for spectroscopic observation as part of the main galaxy sample (MGS; Strauss et al. 2002), which contains all galaxies with $r_{pet} < 17.77$, where $r_{pet}$ is the extinction-corrected *r*-band Petrosian magnitude.

We obtain the SDSS DR7 MGS from the NYU value-added catalogue (NYU VAGC, Blanton et al. 2005a; Padmanabhan et al. 2008; Adelman-McCarthy et al. 2008). The NYU VAGC includes K-corrected absolute magnitudes, and detailed information on the mask. This dataset uses galaxies with $14.5 < r_{pet} < 17.6$. The lower limit ($r_{pet} > 14.5$) guarantees that only galaxies with reliable SDSS photometry are included and the upper limit ($r_{pet} < 17.6$) gives a homogeneous selection over the full footprint of 6141 deg$^2$ (Blanton et al. 2005a). For galaxies that did not obtain a redshift due to fibre collisions, we assign their redshift according to the nearest neighbor.

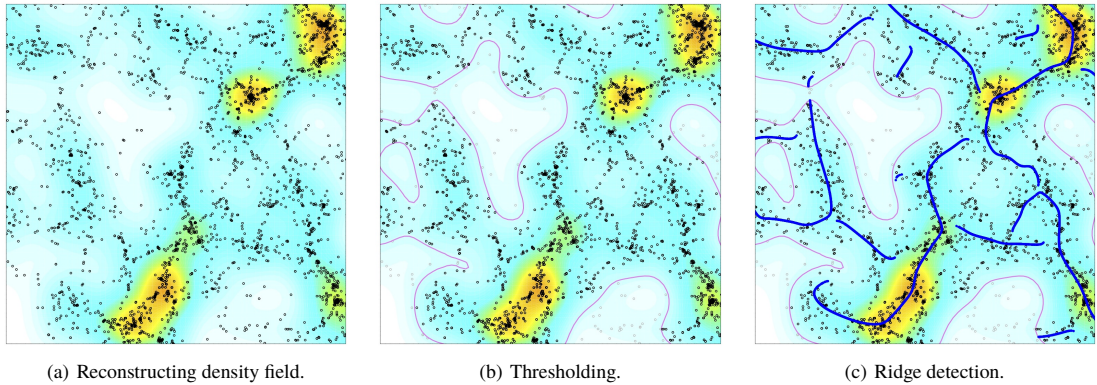

(a) Reconstructing density field.   (b) Thresholding.   (c) Ridge detection.

**Figure 1.** An example of identifying filaments using the density ridge model. Our filament catalogue uses the density ridge model to trace filaments. The catalogue is obtained by applying SCMS algorithm, which consists of three steps: reconstructing density (panel (**a**)), thresholding low-density regions (panel (**b**)), and detecting density ridges (blue curves in panel (**c**)).

The LOWZ and CMASS sample are from Data Release 12 (Alam et al. 2015), the final data of SDSS-III. The Baryon Oscillation Spectroscopic Survey (BOSS) of SDSS-III has obtained spectra and redshifts for around 1.35 million galaxies over a region covering 10,000 square degrees of the sky. These galaxies were selected based on the SDSS imaging (Aihara et al. 2011) and observed along with 160,000 quasars and approximately 100,000 ancillary targets. The targets are assigned to tiles of diameter 3 degrees using the algorithm in Blanton et al. (2003) that adopts to the target density on the sky (Blanton et al. 2003). Spectra are obtained from the BOSS spectrographs (Smee et al. 2013). Each observation is then applied a series of 15-minute exposures, integrating signals until a signal-to-noise ratio threshold is passed for the faint galaxies. This approach yields a homogeneous data set with a high redshift completeness (more than 97%) over the full survey footprint. Finally, redshifts are extracted from the spectra using the approach described in Bolton et al. (2012). A summary of the survey design can be found in Eisenstein et al. (2011); a full description of BOSS is in Dawson et al. (2013).

BOSS selects two different classes of galaxies for spectroscopy : 'LOWZ' and 'CMASS'; detailed description for these two classes can be found in Anderson et al. 2014a. For the LOWZ sample, the effective redshift is $z_{\rm eff} = 0.32$, as we apply a redshift cut $z < 0.43$. The CMASS sample has a median redshift $z = 0.57$ and a stellar mass distribution with maximum at $\log_{10}(M/M_\odot) = 11.3$ (Maraston et al. 2013). The majority of CMASS sample are central galaxies located in dark matter haloes of mass $\sim 10^{13} h^{-1} M_\odot$.

properties of galaxies around filaments[1]. Below we briefly summarize its construction.

The filament catalogue consists of filaments within 130 slices of the Universe from redshift $z = 0.050$ to $z = 0.700$, with slice width $\Delta z = 0.005$ (Chen et al. 2015a). Filaments in each slice are obtained by applying a three-stage procedure (Chen et al. 2015c,a), as explained in the follows.

For each slice, we first smooth galaxies within this slice into a density field. This procedure is done by the kernel density estimator (KDE); namely, the density is given by

$$p(x) = \frac{1}{nb^2} \sum_{i=1}^{N} K\left(\frac{\|x - X_i\|}{b}\right), \qquad (1)$$

where $X_i$ is the $(\alpha_{2000}, \delta_{2000})$ coordinate for $i$-th galaxy within this slice, and $N$ is the total number of galaxies in this slice, and $b$ is the smoothing parameter selected using the rule described in Chen et al. (2015a), which is about 1.5 degree in each redshift slice. We use degree as the smoothing size since the galaxy number density changes drastically from redshift to redshift. The selection of kernel size is still an unsolved problem in statistics and the rule we are applying is at least stable under some toy examples.

After reconstructing the density field, we remove galaxies whose density value is below $\lambda = p_{\rm rms}$. Lastly, we apply the subspace constrained mean shift algorithm (SCMS Ozertem & Erdogmus 2011) based on galaxies pass the density threshold $\lambda$ to obtain filaments. The SCMS detects filaments as ridges (Eberly 1996; Genovese et al. 2014; Chen et al. 2014, 2015c) of the density function in equation (1). Figure 1 provides an illustration of the above process used to construct filaments from given galaxies' positions. More detailed implementation for the filament detection algorithm can be found in Chen et al. (2015a).

By applying the above procedure to each of the 130 slices, we construct a filament catalogue ranging from $z = 0.050$ to $z = 0.700$. We provide examples for detected filaments in the range of the MGS, the LOWZ and the CMASS sample in Figure 2.

## 3 THE FILAMENT CATALOGUE

In this paper, we apply the 'Cosmic Web Reconstruction' catalogue (Chen et al. 2015a), a publicly available filament catalogue, to study

---

[1] The catalogue can be downloaded from https://sites.google.com/site/yenchicr/catalogue.

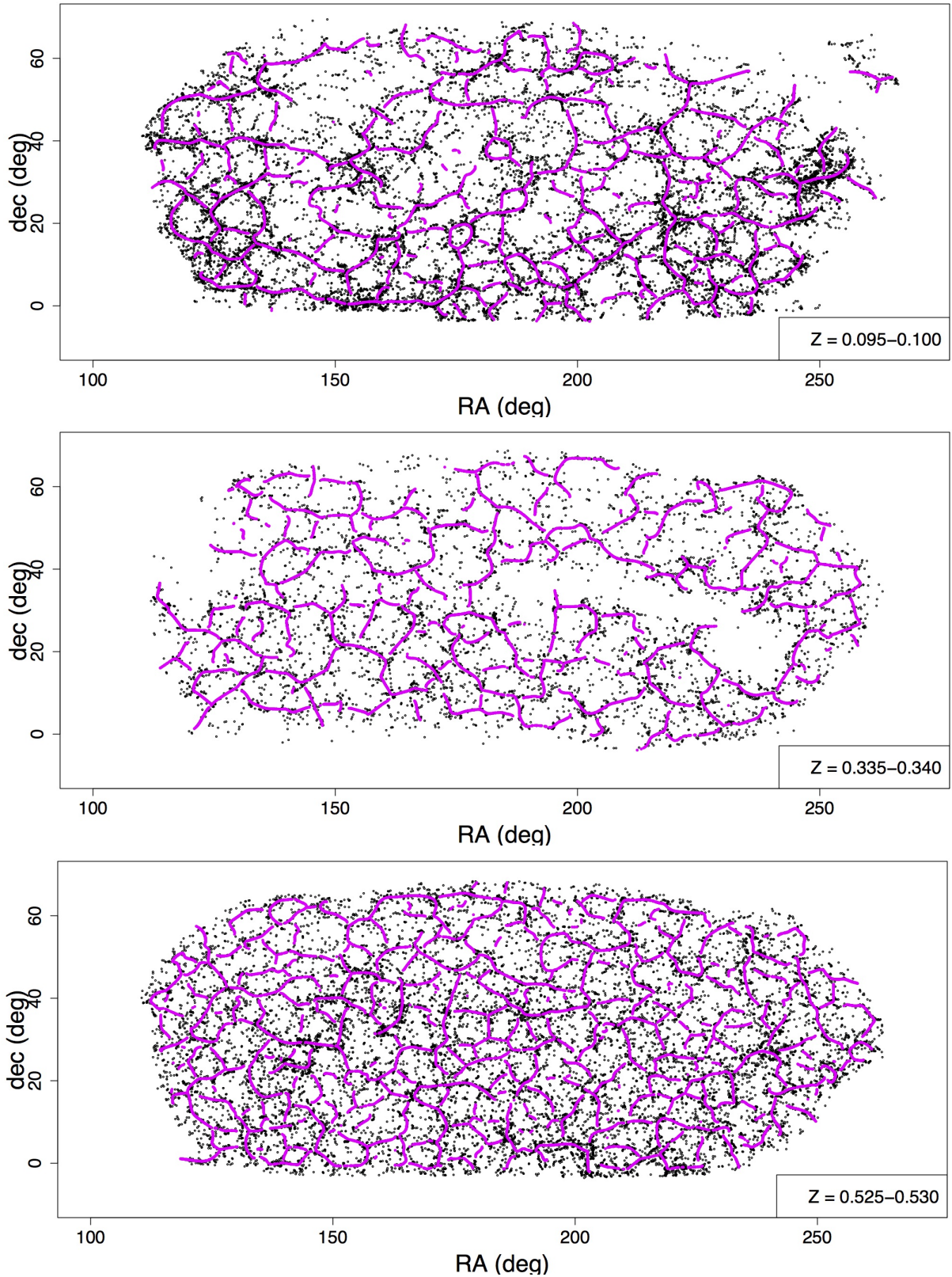

**Figure 2.** Examples of the filament catalogue in narrow redshift slices within the MGS, LOWZ, and CMASS sample, respectively.

## 3.1 Selection of Galaxies for the Study of Effects

To independently analyze the effects from clusters and filaments, we impose two distance cuts to select galaxies. When studying the effects from clusters, we only use galaxies that are 'close to clusters'. When investigating the effects from filaments, we focus on galaxies that are 'away from clusters but close to filaments'.

In more details, when we analyze the effects from clusters, we only consider those galaxies whose *distance to the nearest cluster is less than $R_C$* where

$$R_c = \begin{cases} 20 \text{ Mpc} & \text{for MGS galaxies} \\ 2.5 \text{ Mpc} & \text{for LOWZ galaxies} \\ 10 \text{ Mpc} & \text{for CMASS galaxies} \end{cases} \quad (2)$$

The distance cut $R_c$ is from the study of cluster effect on stellar mass of galaxies; see Appendix A. When we analyze the effect from filaments, we focus galaxies whose *distance to the nearest cluster is at least $R_c$* to remove the effect from clusters and *distance to the nearest filament' is less than $R_f$*, where

$$R_f = \begin{cases} 10 \text{ Mpc} & \text{for MGS galaxies} \\ 30 \text{ Mpc} & \text{for LOWZ galaxies} \\ 40 \text{ Mpc} & \text{for CMASS galaxies} \end{cases} \quad (3)$$

We choose this distance cut $R_f$ for two reasons. First, $R_f$ is about 2 times the average uncertainty in filament position in each sample (Chen et al. 2015a). This choice of $R_f$ behaves like a $2\sigma$ radius. The other reason is based on the observation of density profile (see Appendix B). $R_f$ is roughly the distance that the galaxy number density drops to 50% of the peak number density.

Note that despite the distance cuts $R_c$ and $R_f$ are very large compared to the galactic scale, they are at a similar order of average separate in the absence of clustering in the SDSS. The number density for each sample is about

$$\bar{n} = \begin{cases} 2.3 \times 10^{-3} \text{ Mpc}^{-3} & \text{for the MGS sample} \\ 1.2 \times 10^{-4} \text{ Mpc}^{-3} & \text{for the LOWZ sample} \\ 9.7 \times 10^{-5} \text{ Mpc}^{-3} & \text{for the CMASS sample} \end{cases}, \quad (4)$$

which corresponds to the average separation between two galaxies

$$\frac{1}{\bar{n}^{\frac{1}{3}}} \approx \begin{cases} 7.6 \text{ Mpc} & \text{for MGS galaxies} \\ 20.1 \text{ Mpc} & \text{for LOWZ galaxies} \\ 21.7 \text{ Mpc} & \text{for CMASS galaxies} \end{cases}. \quad (5)$$

Due to the large average separation between two galaxies in the LOWZ and the CMASS sample (this is not a physical fact but is an observational limit), the uncertainty of filaments in the LOWZ and CMASS sample is at a reasonable scale (the filament uncertainty is about $15 - 20$ Mpc at the LOWZ and CMASS sample).

## 4 RESULTS: COLOR

Observational evidence suggests that the color of a galaxy is generally dependent on the environment in which this galaxy resides (Springel et al. 2005; Coil et al. 2008; Cowan & Ivezić 2008; Font et al. 2008; Guo et al. 2011; Grützbauch et al. 2011). If filaments from density ridges trace the high-density environments well, there should be more red galaxies around or in filaments than blue galaxies.

To define red and blue galaxies, we use a simple cut such that a galaxy is classified as a red galaxy if $(g - r) > 0.8$ and is a blue galaxy otherwise. The k-correction for the MGS sample is used to correct to a standard redshift of $z = 0.1$ (note the the k-correction for the LOWZ and CMASS sample is to $z = 0.55$). The first panel in Figure 3 shows the color cut in the color-magnitude diagram. In this comparison, we only use the MGS sample since most galaxies in the LOWZ and CMASS sample are red galaxies (blue galaxies were not selected for spectroscopy for the LOWZ and CMASS sample).

We then compare the average distance to filaments (and clusters) from blue and red galaxies at different redshift regions, denoting $d_F$ and $d_C$ as the distances from a galaxy to its nearest filament and cluster, respectively. This distance is based on 2D projection within each redshift slice. The galaxy clusters are taken from redMaPPer cluster catalogue version 5.10 (Rykoff et al. 2014; Rozo & Rykoff 2014; Rozo et al. 2015). Note that our selection of galaxies by equation (2) and (3) indicates that in all the analysis in this paper, we use galaxies satisfying

$$d_C < R_c \quad (6)$$

for analyzing the effect from clusters and we focus on galaxies with

$$d_C > R_c, \ d_F < R_f \quad (7)$$

for investigating the effect from filaments.

In particular, we compare the scaled distance to filaments (and clusters). The scaled distance is obtained by dividing the distance to filaments (and clusters) by the average distance from all galaxies, i.e., we compare

$$d_F/\langle d_F \rangle_{\text{total}}, \ d_C/\langle d_C \rangle_{\text{total}} \quad (8)$$

from both blue and red populations. The two quantities

$$\langle d_F \rangle_{\text{total}}, \ \langle d_C \rangle_{\text{total}} \quad (9)$$

are the average distance to filaments and clusters for a given subsample by the average distance from all galaxies within the same redshift slice. We include the definitions for the above quantities in Table 1 for reference. We normalize the distance for two reasons. We are only interested in detecting if two populations have different average distances. Scaling the distance by the average over all galaxies does not change the significance of differences between the populations. Due to the change in the number density with redshift in the SDSS, the average distance between a pair of galaxies and the distance to filaments or clusters from a galaxy are changing (the average distance to filaments at different redshifts can be found in Chen et al. 2015a). Without correcting for this effect, the average distance as a function of redshift is difficult to interpret. Scaling the distance by the average over all galaxies eliminates this problem.

The center and right panels of Figure 3 reveal a clear pattern that the red and blue galaxies have significantly different average distances to both filaments and clusters (this separation has a $33.5\sigma$ significance; see Table 2); the red galaxies are closer to both filaments and clusters compared to blue galaxies. And the clusters have a much stronger effect compare to the effect from filaments. Our result is qualitatively consistent with the existing literature; see, e.g., Hogg et al. (2003); Cowan & Ivezić (2008); Grützbauch et al. (2011). Note that we cannot make any interpretation for redshift dependence of the effect from filaments because the uncertainty of locations for filaments increases drastically when the redshift increases. This uncertainty is due to the observational limit and this uncertainty has a much stronger effect than the redshift evolution. So we cannot conclude any redshift dependence based on the current observations.

The significance for comparing average distance to filaments from red and blue galaxies is computed as follows. For each slice in the MGS sample, say slice $\ell$, let $\langle d_F \rangle_{\text{red},\ell}$ and $\langle d_F \rangle_{\text{blue},\ell}$ denote the average distance (without scaling) to filaments from red and blue populations. The quantities $\sigma_{\text{red},\ell}$ and $\sigma_{\text{blue},\ell}$ are the standard errors for $\langle d_F \rangle_{\text{red},\ell}$ and $\langle d_F \rangle_{\text{blue},\ell}$ respectively (standard errors are computed using variance of sample average). Let $S_{\text{MGS}}$ represent the number of slices in the MGS sample (in our case, $S_{\text{MGS}} = 30$). We use the statistic

$$T_{\text{color,MGS}} = \frac{1}{\sqrt{S_{\text{MGS}}}} \sum_{\ell=1}^{S_{\text{MGS}}} \frac{\langle d_F \rangle_{\text{blue},\ell} - \langle d_F \rangle_{\text{red},\ell}}{\sqrt{\sigma_{\text{red},\ell}^2 + \sigma_{\text{blue},\ell}^2}} \qquad (10)$$

to measure the significance for the difference in red and blue populations. If the color is not correlated with distance to filaments, the distribution for the distance from both red and blue populations will be the same, so that $T_{\text{color,MGS}}$ follows a standard normal distribution (mean 0, variance 1) asymptotically. Our computation shows a $58\sigma$ significance for comparing red and blue galaxies. The statistic in equation 10 stacks signals from each slice to enhance the overall signal so that the significance is strong. We also run a KS test and the resulting p-value is similar to the above method.

**Table 1.** Definition of Parameters.

| Parameters | Definition | Comment |
| --- | --- | --- |
| $d_F$ | Distance to the nearest filament. | |
| $d_C$ | Distance to the nearest cluster. | |
| $\langle d_F \rangle_{\text{total}}$ | Average $d_F$ within each redshift slice. | Figure 3 and 4. |
| $\langle d_C \rangle_{\text{total}}$ | Average $d_C$ within each redshift slice. | Figure 3 and 4. |
| $\langle d_F \rangle_{\text{mass,total}}$ | Average $d_F$ within each redshift slice and each fine mass bin. | Figure 6 and 7. |
| $\langle d_C \rangle_{\text{mass,total}}$ | Average $d_C$ within each redshift slice and each fine mass bin. | Figure 6 and 7. |

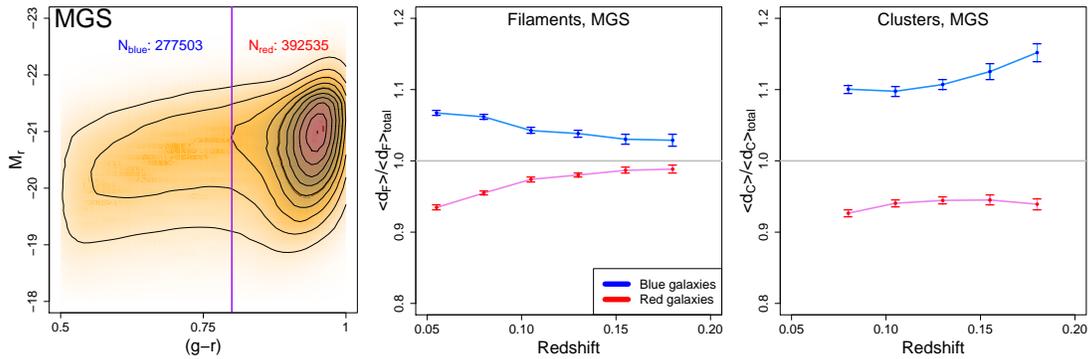

**Figure 3.** The difference between the proximity to filaments (and clusters) from red and blue galaxies in the MGS sample with $z \leqslant 0.2$. **Left panel**: the color-magnitude diagram. The purple line indicates the color cut we use for separating blue (total number $N_{blue}$ = 277, 503) and red galaxies (total number $N_{red}$ = 392, 536). **Center panel**: Scaled distance from red and blue galaxies to filaments. **Right panel**: Scaled distance from red and blue galaxies to clusters. In both center and right panels, blue curves are significantly above red curves, indicating that blue galaxies, on average, have a larger distance to both filaments and clusters than red galaxies.

## 5 RESULTS: STELLAR MASS

In galaxy evolution, one of the most important properties is the stellar mass of a galaxy. Grützbauch et al. (2011) and Alpaslan et al. (2015) found that the stellar mass of a galaxy depends on the environment. Since filaments are tracers of the overdense regions, we expect to see the average distance to filaments from galaxies to be different when galaxies are partitioned according to their stellar mass.

To obtain the stellar mass for each SDSS galaxy, we use the result from the Granada Group. The Granada Group applies the Flexible Stellar Population Synthesis (FSPS) code (Conroy et al. 2009) to SDSS DR12, computing the stellar mass and age for each galaxy[2].

The panels in the first column of Figure 4 display the distribution of stellar mass at different redshifts for each catalogue. We divide galaxies into three mass bins: low-mass, moderate-mass, and high-mass galaxies. Since the mass distribution is different from each catalogue, we use different mass cuts for different catalogues. The mass cuts are chosen to balance the number of galaxies within each bin. The two horizontal black lines in each panel of the first column panels of Figure 4 are the mass cuts.

For each mass bin and each catalogue, we perform the same analysis as in analyzing the color: we first scale the distance within each slice by the average distance from all galaxies and then compute the distance from different galaxies in different mass bin. For the MGS sample, we only consider $z \leqslant 0.20$; for the LOWZ sample, we focus on $0.20 \leqslant z \leqslant 0.43$; for the CMASS sample, we include $0.43 \leqslant z$. The two vertical dashed lines in each panel indicate the boundaries of MGS, LOWZ and CMASS. The two redshift cuts ($z = 0.20, 0.43$) are based on the number density for each catalogue; see Anderson et al. (2014b).

---
[2] More details can be found in http://www.sdss.org/dr12/spectro/galaxy_granada/

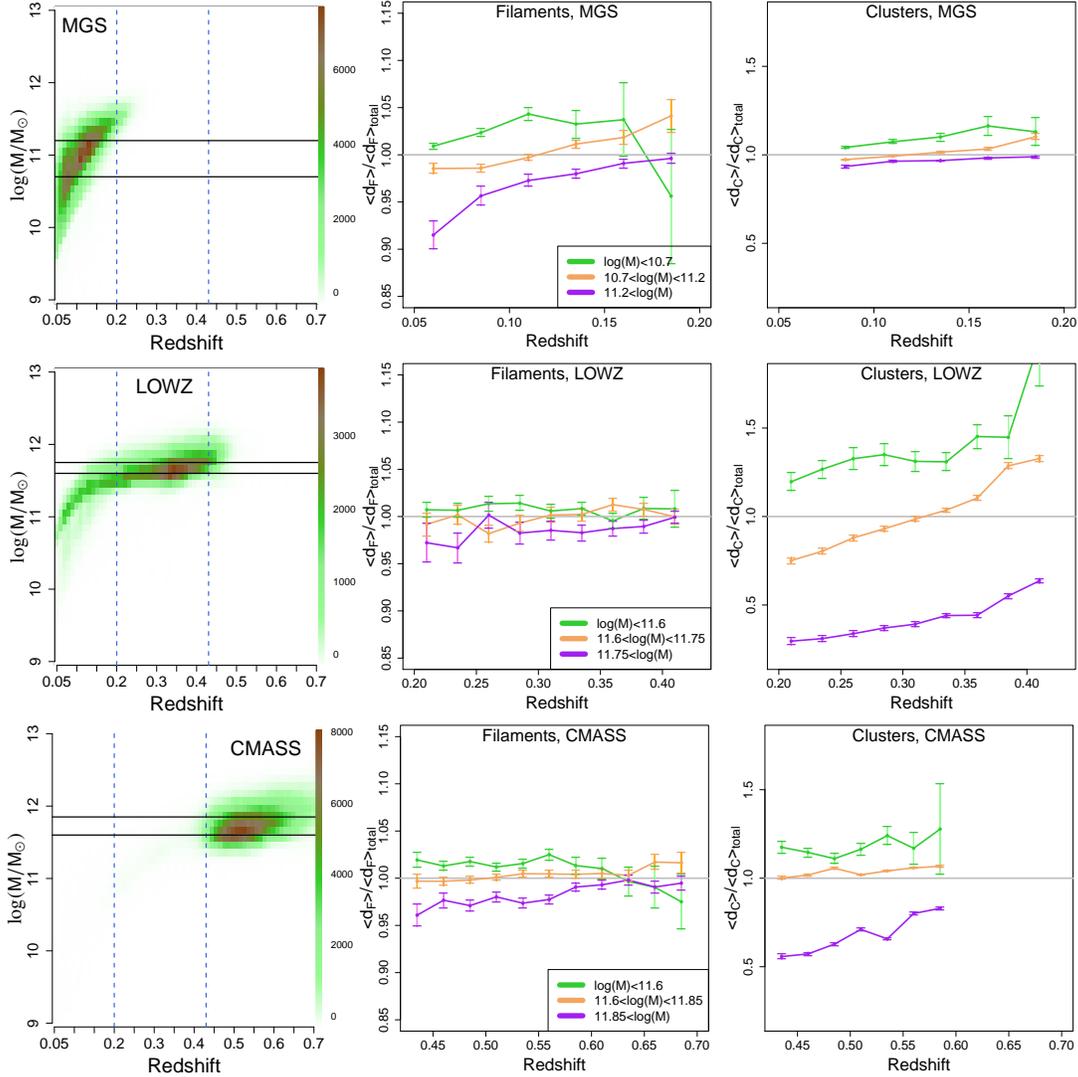

**Figure 4.** The correlation between a galaxy's stellar mass and its proximity to filaments (and clusters). We separate galaxies according to their mass into three bins: low-mass galaxies (green color), moderate-mass galaxies (brown color), and high-mass galaxies (purple color). The top row of panels is the result for the MGS sample; the middle row is for the LOWZ sample; the bottom row is for the CMASS sample. **Left column**: the distribution of the logarithm of the stellar mass for galaxies. The two horizontal black lines indicate the boundaries for each mass bin. The two dashed vertical lines represent the boundaries for the three SDSS galaxy samples. **Center column**: Scaled distance from galaxies in different mass bins to filaments. **Right column**: Scaled distance from galaxies in different mass bins to clusters. There is a consistent trend in all panels of center and right columns: purple curves are lowest and green curves are highest. This means that heavy galaxies are closer to both filaments and clusters on average than light galaxies.

The center and middle columns of Figure 4 display the results. For the proximity to filaments (center column), the average scaled distances in the three mass bins significantly differ from one another. We observe significances $> 2.3\sigma$ using the same formula as in equation (10), indicating that high-mass galaxies tend to appear around filaments than the low-mass galaxies, i.e., a galaxy's stellar mass is correlated with its proximity to filaments.

The right column of Figure 4 presents the same analysis for clusters. We find a clear pattern: the three mass bins have different distances to clusters on average, indicating that stellar mass is correlated with the proximity to clusters. Comparing the result for clusters to filaments demonstrates that in all three samples, the effects from cluster is much stronger than the effect from filaments (note that the scale of Y-axis in the middle and right columns of Figure 4 is different). This result is reasonable since clusters are tracers of the extremely dense regions, while filaments are tracers of regions with a mild overdensity compared to clusters. Since galaxies that reside in high-density regions tend to have more stellar mass (Grützbauch et al. 2011), we expect the high-mass galaxies have a higher change to appear around clusters than low-mass galaxies.

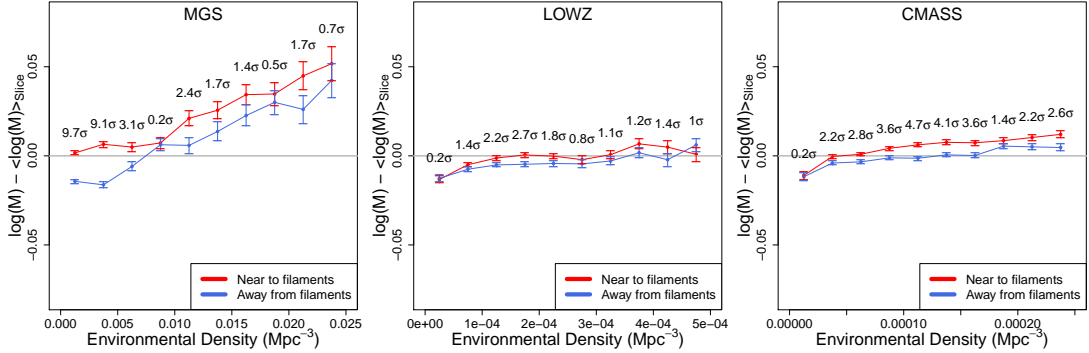

**Figure 5.** The effect from filaments other than the environments. We separate galaxies into to groups, galaxies near to filaments and galaxies away from filaments, based on the median distance to filaments at each slice (so the two groups have equal number of galaxies). Then we plot the scaled stellar mass as a function of the environmental density. We scale the stellar mass by subtracting it by the average of each slice because there is a strong redshift dependence on the mass distribution; see first column of Figure 4. Clearly, under the same environmental density, galaxies near to filaments (red curves) are significantly more massive than galaxies away from filaments (blue curves).

### 5.1 The Effect from Filaments other than Environments

To demonstrate that filaments have an additional effect compared to the effect from the environment, we plot the (scaled) stellar mass as a function of environmental density separately for galaxies near to and away from filament in Figure 5. The environmental density is computed by making boxes on the sky and counts number of galaxies within each box (histograms). The details can be found in Appendix B. We scaled the stellar mass by subtracting it by the average mass within each redshift slice to remove the redshift dependence of the stellar mass (c.f. the first column of Figure 4). We separate galaxies into two groups, near to filaments (red curves) and away from filaments (blues), based on the median distance to filaments. Thus, both groups have equal number of galaxies. We then plot the (scaled) stellar mass as a function of environmental density (galaxy number density). Observing from Figure 5, we find that under the same environmental density, galaxies near to filaments tend to be more massive compare to than those away from filaments. Since we only consider galaxies that are at least $R_c$ away from clusters, this effect is not from clusters. Therefore, Figure 5 provides an evidence for the effect from filaments on galaxy's stellar mass.

### 6 RESULTS: AGE

Besides color and stellar mass, the age of a galaxy is also found to be environment-dependent; see, e.g., (Bernardi et al. 1998; Trager et al. 2000; Wegner & Grogin 2008; Smith et al. 2012a). Generally, early-forming galaxies tend to reside in overdense regions (Sil'chenko 2006; Bernardi et al. 2006; Deng 2014). Based on this density-dependent relation to the age, we expect that early-forming galaxies will be closer to filaments than late-forming galaxies.

To assign an age to each SDSS galaxy, we use the best-fit mass-weighted average age of the stellar population of the FSPS method (Conroy et al. 2009). The age distribution within each galaxy catalogue is given in the first column of Figure 6. There are many strips in the age distribution; this pattern arises because the FSPS method uses a grid of models, where age is a discrete variable.

Since a galaxy's age and stellar mass are correlated, we partition galaxies according to their stellar mass within the same redshift slice to adjust the effect from stellar mass. For the MGS, we construct 30 even log-stellar mass bins ranging from 10.0 to 11.5 (in the unit of logarithm of solar mass and each bin has $\delta \log(M) = 0.05$); the LOWZ sample use 10 bins for the mass ranging from 11.5 to 12.0; the CMASS has 16 bins for the mass ranging from 11.5 to 12.3. Then we compute the average distance to filaments and clusters using all galaxies within the same mass bin and redshift slice. We denote these mass-adjusted average distance as $\langle d_F \rangle_{\text{mass,total}}$ and $\langle d_C \rangle_{\text{mass,total}}$; these quantities are used to normalize $d_F$ and $d_C$ for each galaxies within the specified mass and redshift range as equation (8).

After computing the scaled distance $d_F/\langle d_F \rangle_{\text{mass,total}}$ and $d_C/\langle d_C \rangle_{\text{mass,total}}$, we partition galaxies in each catalogue into three age-types: early-forming galaxies (dark green), intermediate-stage galaxies (orange), and late-forming galaxies (light blue). Similar to the mass cuts, we choose the age cuts to balance the number of galaxies within each type. We conduct the same analysis as for color and stellar mass that compares scaled distance from each age-type of galaxy to filaments and clusters in each catalogue.

For the scaled distance to filaments (center column of Figure 6), every difference between results for galaxies with different formation times in the MGS is significant ($> 6\sigma$ by equation (10); see Table 2). We observe that early-forming galaxies tend to be closer to filaments compared to late-forming galaxies. When we compare to the case of clusters, a similar pattern is observed and clusters have a much stronger signal (note that the scale of Y-axis for cluster cases is different from that of filament cases). Thus, the data suggest that there is correlation between a galaxy's age and its proximity to filaments (and clusters) after adjusting the effect from stellar mass. A possible scenario is that early-forming galaxies tend to sit at the centers of halos which will eventually become clusters or form filaments. We have not considered the color-age dependency and quenching; these are possible effects that could explain the age-proximity correlation.

For the LOWZ and the CMASS sample, there is no significant difference between different age groups of galaxies for filament cases. For the cluster cases, in the LOWZ sample we observe that the intermediate age galaxies might be farther to clusters compare to the early-forming galaxies. In the CMASS sample, the separations are insignificant. One possible reason is that the age estimate for galaxies is not accurate for the LOWZ and the CMASS sample.

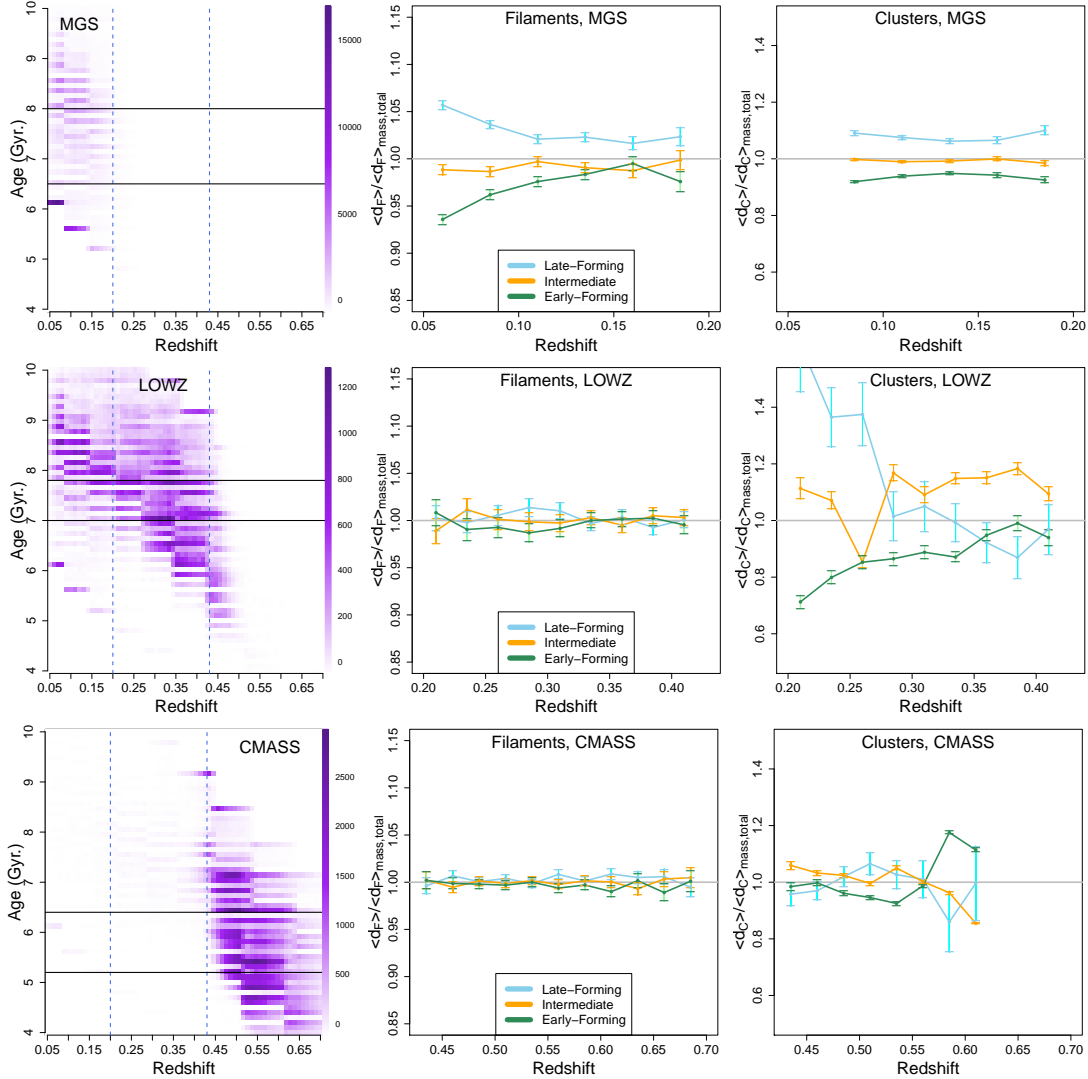

**Figure 6.** The correlation between a galaxy's age and its proximity to filaments (and clusters). We partition galaxies into three age types: late-forming galaxies (light blue), intermediate-stage galaxies (orange), and early-forming galaxies (dark green). The top row of panels is the result for the MGS sample; the middle row is for the LOWZ sample; the bottom row is for the CMASS sample. **Left column**: the distribution of the age for galaxies. The top horizontal black lines indicate the boundaries for each age-type. The two horizontal black lines indicate the boundaries for each age-type. The two dashed vertical lines represent the boundaries for the three SDSS galaxy samples. **Center column**: Scaled distance from different age-type galaxies to filaments. **Right column**: Scaled distance from different age-type galaxies to clusters. On the first row (MGS), there is a clear pattern that old galaxies (green curves) are closer to both the filaments and clusters than young galaxies (light blue curves) after adjusting the effect from stellar mass. However, for the LOWZ and the CMASS sample, we do not observe a significant signal for the age in the filament cases.

## 7 RESULTS: SIZE

Finally, we investigate the relation of the size of a galaxy with its proximity to filaments. Similar to the color and the stellar mass, the size of a galaxy has also been found to be dependent on the environment (Cooper et al. 2012; Lani et al. 2013; Cappellari 2013; Kelkar et al. 2015). Hence, due to the properties of filaments, we expect to see a difference in the average distance to filaments from galaxies when we partition galaxies according to their size.

Our analysis focuses on the LOWZ sample since photometry in the CMASS sample is not as accurate as the LOWZ sample, and faint galaxies in the MGS sample also suffer from this issue. The size adopted is the effective radius ($R_e$) obtained by fitting the de Vaucouleurs profile (de Vaucouleurs 1948). We partition galaxies into three groups by their size:

small galaxies (green) : $R_e < 5.6$ kpc medium galaxies (brown) : $5.6$ kpc $< R_e < 7.8$ kpc large galaxies (purple) : $7.8$ kpc $< R_e$.

We select two thresholds (5.6 and 7.8 kpc) to balance the number of galaxies within each size-type. The size distribution is presented in the left panel of Figure 7. Since stellar mass has influence over the average distance, we apply the mass-partitioning method in previous section to adjust the effect of stellar mass.

The results are presented in the center and right panels of

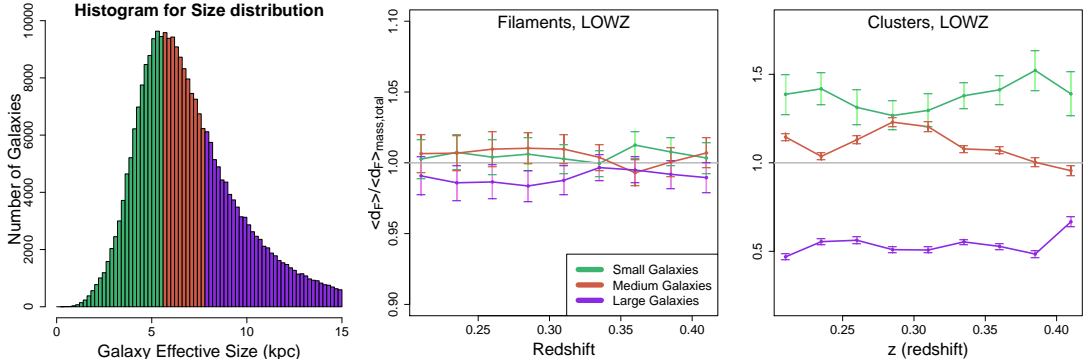

**Figure 7.** The correlation between a galaxy's size and its proximity to filaments (and clusters). We partition galaxies into three types: small galaxies (green, $R_e < 5.6$ kpc), medium galaxies (dark orange, 5.6 kpc $< R_e < 7.8$ kpc), and large galaxies (purple, 7.8 kpc $< R_e$). **Left panel**: the age distribution for galaxies. The histogram is colored according to the three size. **Center panel**: Scaled distance from different size galaxies to filaments. **Right panel**: Scaled distance from different size galaxies to clusters. In both center and right panel, large galaxies (purple curves) are always lowest than other two size-types of galaxies after adjusting the effect from stellar mass, indicating that large galaxies tend to be closer to both filaments and clusters than small galaxies.

Figure 7. There is a significant difference in average distance to filaments and clusters when large galaxies are compared to the other two types (significance $> 2.9\sigma$ by equation (10); see Table 2). This result indicates that, after adjusting the effect from stellar mass, large galaxies (purple) aggregate around filaments and clusters while small galaxies (green) tend to have a larger average distance from both filaments and clusters. We do not observe any significance for the difference between small and medium galaxies in the filament case. Note that in the cluster cases, the separations are very significant.

Our result is consistent with previous studies that galaxies that reside within denser environments tend to be larger (Lani et al. 2013; Cooper et al. 2012). A possible explanation for this effect is galaxy mergers. Mergers are more common in high-density environments so that if mergers are the dominating effect for the size growth of early-type massive galaxies, we should expect to see a correlation between density environment and the size of a galaxy (Cooper et al. 2012). Since filaments and clusters are tracers of high-density regions, a galaxy with a shorter distance to filaments (or clusters) will have a higher density environment than galaxies located at a greater distance.

## 8 DISCUSSION

In this paper, we study the relationship of properties of a galaxy as a function of its distance to the nearest filaments (and clusters). We observe strong separations between different types of galaxies; table 2 summarizes the signal strengths for each catalogue (the MGS, LOWZ, and CMASS). For the MGS, the separations between types are significant among all comparisons. Differences with $> 6.0\sigma$ are present for all comparisons; some relations even have a $> 15\sigma$ significance. For the LOWZ sample, the results for stellar mass are significant ($> 2.3\sigma$). In the size case, after adjusting the effect from stellar mass, we obtain significant correlations for medium galaxies versus large galaxies ($> 2.9\sigma$). In the CMASS sample, separating galaxies by stellar mass yields a significant result ($> 4.1\sigma$). Taking all evidence into account, our analysis provides evidence that several galaxy properties are correlated with filaments. In figure 8, we provide an illustration about how galaxy properties and filaments are correlated using the slice $z = 0.095 - -0.100$. In each panel, we focus on one galaxy property (age, mass, or age) and partition the entire region into $1 \times 1$ deg$^2$ cells. For each cell, we compute the average value for that galaxy property and color each cell based on the within-cell average. Finally, we show filaments using black curves. This gives us a clear picture on how these galaxy properties are correlated with filaments.

Moreover, we also observe from Figure 5 that filaments have additional effects compare to the effects from environments. Under the same environmental densities, galaxies tend to be more massive when they are closer to filaments.

Our findings also include:

(i) Several galaxy properties, including color, stellar mass, age, and size, are correlated with filaments. Our result is direct evidence for the correlation between filaments and the properties of galaxies (other results can be found in Guo et al. 2015; Alpaslan et al. 2015; Eardley et al. 2015). These correlation signals are expected since filaments trace the medium-to-high environmental density regions.

(ii) Even in the high redshift regime ($z > 0.25$), we still detect a consistent correlation signal from many galaxy properties to filaments. Other analysis uses the MGS sample, which focuses on the regions at $z \leqslant 0.25$ (Zhang et al. 2013; Tempel et al. 2014b; Guo et al. 2015). Filament analysis at high redshift ($z \geqslant 0.25$) has been done in some other surveys, see, e.g., Eardley et al. (2015); Alpaslan et al. (2015); our findings are consistent with theirs. These high redshift surveys, however, focus only on a small region of the sky.

(iii) Filaments from Cosmic Web Reconstruction catalogue and reMaP-Per galaxy clusters are similar in the sense that they have similar observed trends in galaxy properties near both filaments and clusters. This behavior shows the effectiveness of using density ridges as tracers of filaments. Prior to our work, no direct analysis of filament impacts on galaxy properties has been performed for the filaments from density ridges, although density ridges are similar to Voronoi filaments (Chen et al. 2015c) and have the desired statistical properties (Chen et al. 2014).

(iv) The cluster effect on the stellar mass of a galaxy is sensitive to the filamentary environments this galaxy resides in. Galaxies that are close to both a filament and a cluster tend to be more massive than galaxies that are only close to a cluster. This reveals that filaments and galaxy clusters have distinct effects on galaxy properties.

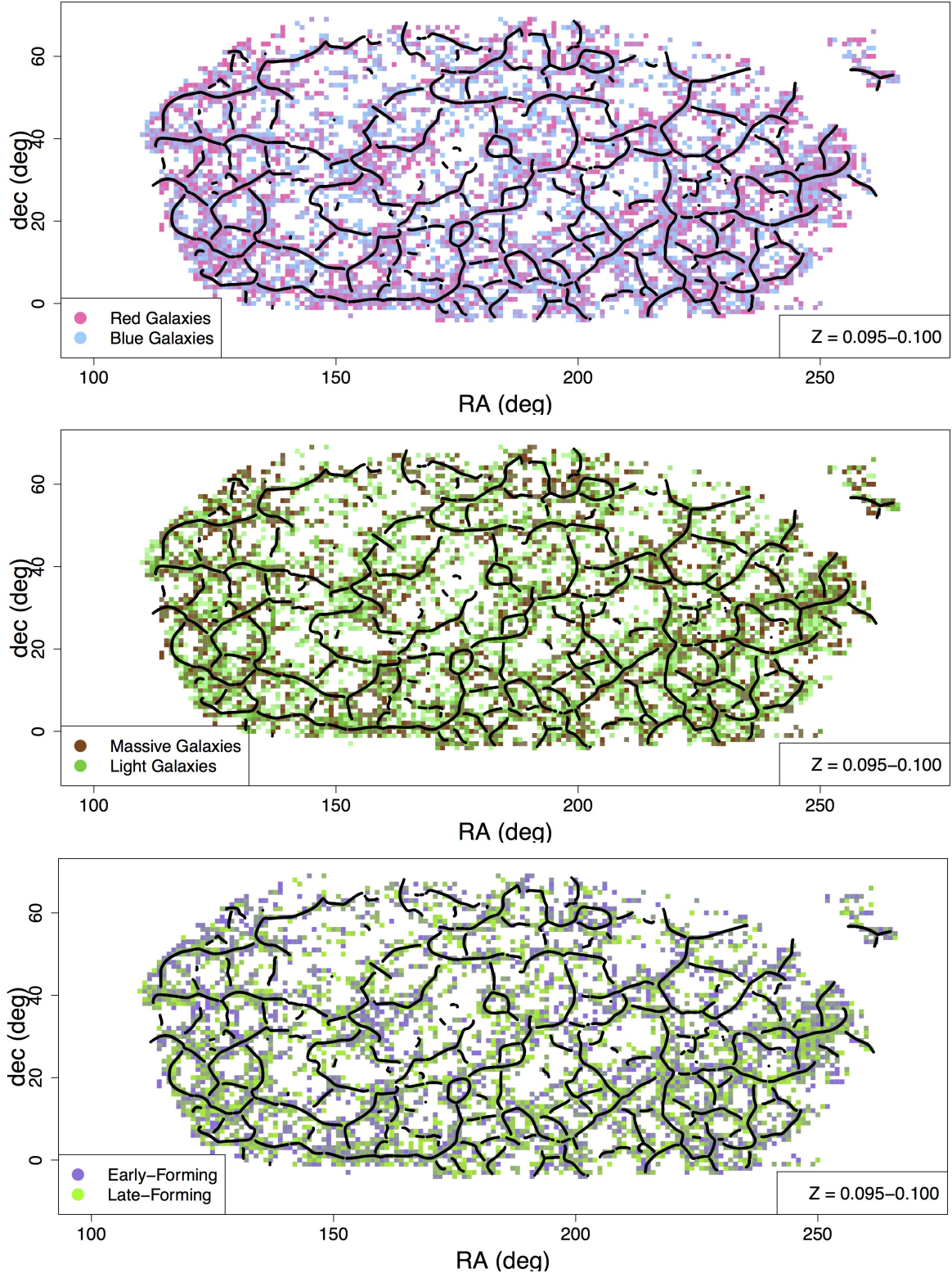

**Figure 8.** An example for the filament effect on galaxy properties. This is the slice of $z = 0.095 - 0.100$. We partition the entire region into $1 \times 1$ deg$^2$ cells and compute the average value for each of the galaxy property within each cell. The black curves are filaments. **Top:** we show the average color (by $(g - r)$ value) within each cell versus filaments. **Middle:** we show the average mass within each cell versus filaments. **Bottom:** we show the average age within each cell versus filaments. This visual comparison gives us a clear picture about how filaments and galaxy properties are correlated.

| Separation | Type | MGS ($0.05 < z < 0.20$) | LOWZ ($0.20 < z < 0.43$) | CMASS ($0.43 < z < 0.70$) |
|---|---|---|---|---|
| Color | Red VS Blue | $33.5\sigma$ | N/A | N/A |
| Stellar Mass | Low-mass VS Moderate-mass | $9.6\sigma$ | $2.3\sigma$ | $4.1\sigma$ |
| | Moderate-mass VS High-mass | $10.0\sigma$ | $3.4\sigma$ | $9.4\sigma$ |
| Age | Intermediate-stage VS Early-forming | $6.0\sigma$ | $0.9\sigma$ | $1.3\sigma$ |
| | Late-forming VS Intermediate-stage | $11.9\sigma$ | $0.4\sigma$ | $1.2\sigma$ |
| Size | Small VS Medium | N/A | $0.1\sigma$ | N/A |
| | Medium VS Large | N/A | $2.9\sigma$ | N/A |

**Table 2.** Signal strength for the separation of galaxy distance to filaments according to the color, stellar mass, age, and size.

## ACKNOWLEDGMENTS

We thank Hung-Jin Huang, Florent Leclercq, Peter Melchior, Dmitri Novikov, and Hy Trac for useful comments and discussions; we also thank Sukhdeep Singh for providing the de Vaucouleurs sizes for the LOWZ sample. This work is supported in part by the Department of Energy under grant DESC0011114; SH is supported in part by DOE-ASC, NASA and NSF; RM is supported by the Alfred P. Sloan Foundation; CG is supported in part by DOE and NSF; LW is supported by NSF grant DMS1513412.

Funding for SDSS-III has been provided by the Alfred P. Sloan Foundation, the Participating Institutions, the National Science Foundation, and the U.S. Department of Energy Office of Science. The SDSS-III web site is http://www.sdss3.org/.

SDSS-III is managed by the Astrophysical Research Consortium for the Participating Institutions of the SDSS-III Collaboration including the University of Arizona, the Brazilian Participation Group, Brookhaven National Laboratory, Carnegie Mellon University, University of Florida, the French Participation Group, the German Participation Group, Harvard University, the Instituto de Astrofisica de Canarias, the Michigan State/Notre Dame/JINA Participation Group, Johns Hopkins University, Lawrence Berkeley National Laboratory, Max Planck Institute for Astrophysics, Max Planck Institute for Extraterrestrial Physics, New Mexico State University, New York University, Ohio State University, Pennsylvania State University, University of Portsmouth, Princeton University, the Spanish Participation Group, University of Tokyo, University of Utah, Vanderbilt University, University of Virginia, University of Washington, and Yale University.

## APPENDIX A: THE EFFECT OF RADIUS FROM CLUSTERS

We study the effect of radius from clusters using galaxy's stellar mass. We first scale the stellar mass by the average of each redshift slice to remove the redshift dependence of stellar mass (c.f. first column of Figure 4). Then we plot the average (scaled) mass as a function of distance to clusters. From Figure A1, we see that the stellar mass decreases rapidly when the distance to cluster increases. For MGS galaxies, clusters have significant effect ($> 2\sigma$) on stellar mass up to $R_c$(MGS) = 20 Mpc; for LOWZ galaxies, the effect of radius $R_c$(LOWZ) is only 2.5 Mpc; for CMASS galaxies, we have an effect of radius $R_c$(CMASS) = 10 Mpc. The difference of effect of radius might be due to the quality of galaxy clusters in redMaPPer at different redshift range. RedMaPPer clusters are robust under redshift $z \in [0.1, 0.33]$ (Rozo & Rykoff 2014) so that most of the LOWZ sample is covered within this range. For clusters in the MGS and the CMASS sample, the reMaPPer has more missing clusters so that the effect of radius is smoothed out by these missing clusters, which increases the effect of radius.

## APPENDIX B: THE DENSITY PROFILE OF FILAMENTS

In this section, we study the density profile of filaments. The galaxy number density is computed as follows. For each redshift slice, we use a 2 dimensional histogram with window size $2 \times 2$ deg$^2$. We count the number of galaxies within each window and divide the number by the size of each within and the width of the redshift slice $\Delta z = 0.005$ to obtain the estimate of number density. Note that we have also try a $1 \times 1$ deg$^2$ window size and the result remains similar. Finally, we plot the galaxy number density as a function of distance to filaments and the result is given in Figure B1. In the MGS sample, we see a rapid drop in number density. But in both the LOWZ and CMASS sample, the decreasing pattern is slower. This is because we have a much smaller number density in the LOWZ and the CMASS sample. Note that the filament distance cut $R_f$ (see equation (3)) is about the distance for 50% of the peak of filament density profile for each sample.

This paper has been typeset from a T<sub>E</sub>X/LAT<sub>E</sub>X file prepared by the author.

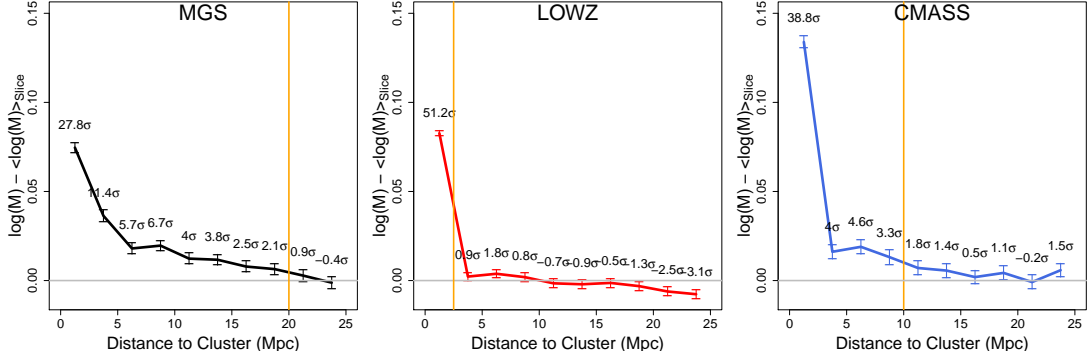

**Figure A1.** The effect of radius of clusters on galaxies' stellar mass. We display how the (scaled) mass decreases as a function of distance to clusters. This shows the effect range from clusters. From these three panels, we found that the cluster effect of radius (significance is less than $2\sigma$) on the stellar mass in the MGS/LOWZ/CMASS sample is 20/2.5/10 Mpc (orange line). Thus, in our analysis for the effect from filaments, we only consider galaxies that are at least 20/2.5/10 Mpc away in the MGS/LOWZ/CMASS sample.

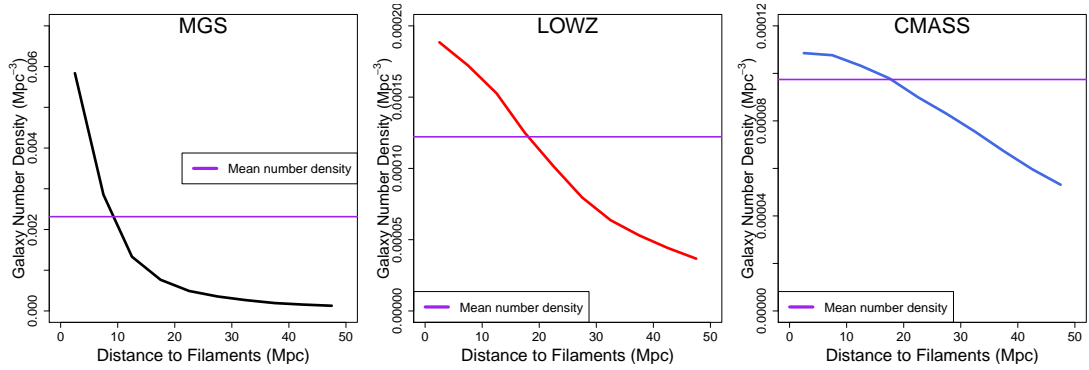

**Figure B1.** The density profile for filaments at the three samples. We plot the galaxy number density as a function of distance to filaments. The purple horizontal line indicates the average number density for each sample. Note that our filament distance cut $R_f$ given in equation (3) is roughly the distance where the number density drops to 50% of the peak number density.